\documentclass[12pt]{book}

\usepackage[dvips]{graphicx,color}
\usepackage{makeidx,tsukuba}
\makeauthorindex
\makeindex

\begin{document}

\BookTitle{\itshape The 28th International Cosmic Ray Conference}
\CopyRight{\copyright 2003 by Universal Academy Press, Inc.}
\pagenumbering{arabic}

\chapter{VHE Observations of BL Lacertae Objects: 1995-2000}

\author{%
D.~Horan,$^{1,2}$ M.A.~Catanese$^1$, I.H.~Bond, P.J.~Boyle,
S.M.~Bradbury, J.H.~Buckley, D.~Carter-Lewis, O.~Celik, W.~Cui,
M.~Daniel, M.~D'Vali, I.de~la~Calle~Perez, C.~Duke, A.~Falcone,
D.J.~Fegan, S.J.~Fegan, J.P.~Finley, L.F.~Fortson, J.~Gaidos,
S.~Gammell, K.~Gibbs, G.H.~Gillanders, J.~Grube, J.~Hall, T.A.~Hall,
D.~Hanna, A.M.~Hillas, J.~Holder, A.~Jarvis, M.~Jordan, G.E.~Kenny,
M.~Kertzman, D.~Kieda, J.~Kildea, J.~Knapp, K.~Kosack, H.~Krawczynski,
F.~Krennrich, M.J.~Lang, S.~LeBohec, E.~Linton, J.~Lloyd-Evans,
A.~Milovanovic, P.~Moriarty, D.~Muller, T.~Nagai, S.~Nolan, R.A.~Ong,
R.~Pallassini, D.~Petry, B.~Power-Mooney, J.~Quinn, M.~Quinn,
K.~Ragan, P.~Rebillot, P.T.~Reynolds, H.J.~Rose, M.~Schroedter,
G.~Sembroski, S.P.~Swordy, A.~Syson, V.V.~Vassiliev, S.P.~Wakely,
G.~Walker, T.C.~Weekes, J.~Zweerink \\ {\it (1) Smithsonian
Astrophysical Obs., P.O. Box 97, Amado, AZ 85645, USA}\\ {\it (2) The
VERITAS Collaboration--see S.P.Wakely's paper} ``The VERITAS
Prototype'' {\it from these proceedings for affiliations} }

\section*{Abstract}
The results of observations of 29 BL Lacertae objects taken with the
Whipple Observatory 10 m gamma-ray Telescope between 1995 and 2000 are
presented. 

\section{Introduction}
Among blazars, BL Lacertae objects (BL Lacs) are believed to be the
best candidates for VHE emission. Consistent with these expectations,
the four confirmed sources of extragalactic gamma rays (E $>$
250\,GeV) are BL Lacs, Mrk\,421 [7], Mrk\,501 [8], H1426 [5] and
1ES1959 [6]. To improve our understanding of the emission mechanisms
in BL Lacs, more need to be detected at very high energies.

\section{Observations and Analysis}

The observations presented here were taken as part of three BL Lac
campaigns. The first was a survey of all known (circa 1995) BL Lacs
with redshift $< 0.1$. The second was a search for TeV emission from
high frequency peaked BL Lacs in the redshift range from 0.1 to
0.2. More recent observations were taken as part of a ``Snapshot
Survey'' in which many BL Lacs were observed for 10 minutes each on a
regular basis in the hope of catching one of them in a flaring state
[3].

Throughout the course of these observations, the imaging camera on the
Whipple telescope was upgraded a number of times, the triggering
criteria were changed and different light concentrators were
installed in front of the camera. Therefore, the peak response energy
of the instrument varied during this period. These observations were
all taken in the ``TRACKING mode'' described in [1].

\begin{table}[t]
 \caption{Observation Results - I}
\begin{center}
\begin{tabular}{lc|ccccccc}
\hline
                            &         &                        & Exp.            &  Total      & Flux             &          &E$_{peak}$$^{c}$\\
Object                      & $z$     & Period$^a$             & [hrs]           &  $\sigma$   & [Crabs]          & Flux$^b$ & [GeV]  \\
\hline
1ES0033+595$^\star$         & 0.086   & 95/12                  &  1.85           & -0.59       &  $<$0.200        & $<$2.10  & 350    \\
1ES0145+138                 & 0.125   & 96/10-96/11            &  7.85           & -1.01       &  $<$0.093        & $<$0.98  & 350    \\
                            &         & 98/11-98/12            &  2.29           &  0.22       &  $<$0.512        & $<$3.50  & 500    \\
                            &         & 98/12-99/01            &  1.98           & -0.50       &  $<$0.357        & $<$3.34  & 500    \\
RGB0214+517$^\star$         & 0.049   & 99/12-00/01            &  6.01           &  0.29       &  $<$0.165        & $<$1.45  & 430    \\
3C66A$^{\dagger,\star,d,e}$ & 0.444   & 95/10-95/11            &  8.00           & -2.00       &  $<$0.056        & $<$0.59  & 350    \\
1ES0229+200$^\star$         & 0.140   & 96/11-96/12            &  7.85           &  0.15       &  $<$0.113        & $<$1.19  & 350    \\
                            &         & 98/11-98/12            &  2.30           & -1.08       &  $<$0.326        & $<$2.23  & 500    \\
                            &         & 98/12-99/01            &  1.78           & -0.40       &  $<$0.403        & $<$3.76  & 500    \\
1H0323+022$^\star$          & 0.147   & 96/11-96/12            & 10.18           &  1.02       &  $<$0.181        & $<$1.90  & 350    \\
                            &         & 97/01                  &  0.91           &  0.20       &  $<$0.298        & $<$3.13  & 350    \\
                            &         & 98/12-99/01            &  3.18           &  1.69       &  $<$0.509        & $<$4.75  & 500    \\
EXO0706.1+5913              & 0.125   & 96/12                  &  5.55           & -1.16       &  $<$0.087        & $<$0.91  & 350    \\
                            &         & 97/01-97/03            &  3.69           &  0.76       &  $<$0.161        & $<$1.69  & 350    \\
                            &         & 98/11                  &  1.83           & -0.40       &  $<$0.524        & $<$3.58  & 500    \\
                            &         & 98/12-99/02            &  1.90           &  0.07       &  $<$0.459        & $<$4.29  & 500    \\
1ES0806+524$^\star$         & 0.138   & 96/02-96/03$^\bullet$  &  5.57           &  0.46       &  $<$0.104        & $<$1.09  & 350    \\
                            &         & 00/01-00/03            &  4.16           & -0.29       &  $<$1.293        & $<$11.4  & 430    \\
PKS0829+046$^{\dagger,e}$   & 0.180   & 95/01-95/04            & 11.07           &  1.25       &  $<$0.117        & $<$1.47  & 300    \\
1ES0927+500                 & 0.188   & 96/12                  &  5.08           & -1.92       &  $<$0.064        & $<$0.67  & 350    \\
                            &         & 97/01-97/04            &  5.04           & -1.03       &  $<$0.076        & $<$0.80  & 350    \\
S40954+658$^{\dagger,e}$    & 0.368   & 95/02-95/03            &  3.70           & -1.09       &  $<$0.096        & $<$1.21  & 300    \\
1ES1028+511$^\star$         & 0.361   & 98/12-99/02            &  4.43           &  0.57       &  $<$0.287        & $<$2.68  & 500    \\
1ES1118+424                 & 0.124   & 98/02-98/04            &  7.30           & -0.25       &  $<$0.218        & $<$1.49  & 500    \\
                            &         & 98/12-99/02            &  3.60           &  0.27       &  $<$0.310        & $<$2.90  & 500    \\
                            &         & 00/01-00/05            &  6.97           & -0.62       &  $<$0.116        & $<$1.02  & 430    \\
Mrk40                       & 0.021   & 00/01-00/04            & 10.16           &  2.59       &  $<$0.206        & $<$1.81  & 430    \\
Mrk180$^\star$              & 0.046   & 95/01-95/04$^\bullet$  &  5.55           & -0.10       &  $<$0.108        & $<$1.36  & 300    \\
                            &         & 95/12-96/05$^\bullet$  & 20.46           & -0.26       &  $<$0.105        & $<$1.10  & 350    \\
                            &         & 97/01$^\bullet$        &  0.79           & -0.17       &  $<$0.303        & $<$3.18  & 350    \\
1ES1212+0748                & 0.130   & 99/02                  &  1.13           &  0.44       &  $<$0.778        & $<$7.26  & 500    \\
                            &         & 00/01-00/05            &  3.70           &  1.30       &  $<$0.321        & $<$2.82  & 430    \\
ON325$^{\dagger,\star}$     & 0.237   & 99/02                  &  0.97           &  1.27       &  $<$0.882        & $<$8.23  & 500    \\
                            &         & 00/01-00/05            &  5.05           &  0.88       &  $<$0.215        & $<$1.89  & 430    \\
1H1219+301$^\star$          & 0.182   & 95/01-95/05            &  2.77           &  2.71       &  $<$0.226        & $<$2.85  & 300    \\
                            &         & 97/02-97/06            & 11.27           &  0.99       &  $<$0.079        & $<$0.83  & 350    \\
                            &         & 98/01-98/03            &  1.38           & -1.96       &  $<$0.356        & $<$2.43  & 500    \\
                            &         & 98/12-99/02            &  2.94           & -0.08       &  $<$0.296        & $<$2.77  & 500    \\
                            &         & 00/01-00/04            &  3.69           &  0.04       &  $<$0.191        & $<$1.68  & 430    \\
\hline
\end{tabular}
\end{center}
\end{table}

\begin{table}[t]
 \caption{Observation Results - II}
\begin{center}
\begin{tabular}{lc|cccccc}
\hline
                            &           &                        & Exp.            &  Total      & Flux             &          &E$_{peak}$$^{c}$  \\
Object                      & $z$       & Period$^a$             & [hrs]           &  $\sigma$   & [Crabs]          & Flux$^b$ & [GeV]\\
\hline
WComae$^{\dagger,e}$        & 0.102     & 95/02-95/04            & 14.33           & -0.57       &  $<$0.052        & $<$0.66  & 300  \\
                            &           & 96/01-96/05            & 15.73           & -0.29       &  $<$0.055        & $<$0.58  & 350  \\
                            &           & 99/01-99/02            &  4.43           & -0.03       &  $<$0.312        & $<$2.92  & 500  \\
                            &           & 00/01-00/04            &  4.72           & -0.58       &  $<$0.148        & $<$1.30  & 430  \\
MS1229.1+6430               & 0.170     & 95/02-95/04            &  1.39           &  1.32       &  $<$0.286        & $<$3.60  & 300  \\
                            &           & 99/02                  &  2.04           & -0.76       &  $<$0.446        & $<$4.16  & 500  \\
                            &           & 00/01-00/05            &  6.01           &  0.35       &  $<$0.170        & $<$1.50  & 430  \\
1ES1239+069                 & 0.150     & 99/01-99/02            &  1.73           &  0.78       &  $<$0.616        & $<$6.04  & 500  \\
                            &           & 00/01-00/05            &  5.08           &  0.11       &  $<$0.197        & $<$1.73  & 430  \\
1ES1255+244                 & 0.141     & 97/02-97/05            &  5.54           &  1.19       &  $<$0.112        & $<$1.18  & 350  \\
                            &           & 98/03                  &  0.46           &  0.13       &  $<$1.112        & $<$7.60  & 500  \\
                            &           & 99/02                  &  1.73           &  0.15       &  $<$0.508        & $<$4.75  & 500  \\
                            &           & 00/01-00/05            &  4.16           & -0.54       &  $<$0.164        & $<$1.45  & 430  \\
OQ530$^*$                   & 0.151     & 95/03-95/05            &  7.39           & -0.73       &  $<$0.058        & $<$0.73  & 300  \\
4U1722+11$^{\star,f}$       & 0.018     & 95/04-95/05$^\bullet$  &  2.77           & -0.08       &  $<$0.124        & $<$1.56  & 300  \\
IZw187$^\star$              & 0.055     & 95/03-95/04$^\bullet$  &  2.31           & -1.27       &  $<$0.086        & $<$1.08  & 300  \\
                            &           & 96/04-96/05$^\bullet$  &  2.32           &  0.61       &  $<$0.150        & $<$1.58  & 350  \\
1ES1741+196$^\star$         & 0.084     & 96/05-96/07$^\bullet$  &  9.23           & -1.02       &  $<$0.053        & $<$0.56  & 350  \\
                            &           & 98/05                  &  0.46           & -0.08       &  $<$1.168        & $<$7.99  & 500  \\
3C371$^*$                   & 0.051     & 95/05-95/06            & 13.04           &  0.41       &  $<$0.190        & $<$1.23  & 300  \\
BLLac$^{\dagger,\star,d,e}$ & 0.069     & 95/07$^\bullet$        &  4.62           &  1.07       &  $<$0.109        & $<$1.37  & 300  \\
                            &           & 95/10-95/11$^\bullet$  & 39.09           & -1.48       &  $<$0.038        & $<$0.40  & 350  \\
                            &           & 98/05-98/06            &  0.92           &  0.47       &  $<$1.722        & $<$8.02  & 500  \\
1ES2321+419                 & 0.059     & 95/10-95/11            &  6.42           & -1.07       &  $<$0.101        & $<$1.06  & 350  \\
\hline
\end{tabular}
\end{center}
$^\dagger$ Low frequency peaked BL Lacs; all others are high frequency peaked BL Lacs.\\
$^\star$ Included in the list of Costamante et al. [2] as a possible TeV emitter.\\
$^a$ Seasons during which the flux upper limit from this
object was found to be lower than that predicted in [2] adapting the model of [4] are marked with a $^\bullet$.\\
$^b$ The integral flux upper limits are quoted above the peak
response energy for the observation period in units of 10$^{-11}$
cm$^{-2}$ s$^{-1}$.\\
$^c$ The peak response energy is the energy at which the collection
area folded with an E$^{-2.5}$ spectrum reaches a maximum. The fact
that it has increased somewhat over time simply reflects the fact that
the energy at which the telescope is most efficient at detecting gamma
rays has increased; e.g., the collection area at 300 GeV in the 1999-2000 
season, was greater than that at this energy in 1995.\\
$^d$Unconfirmed source of TeV gamma rays.\\
$^e$EGRET source of $>$100 MeV gamma rays.\\
$^f$This redshift estimate is based on just one absorption line [9].\\

\end{table}

\section{Results And Discussion}

Tables 1 \& 2 summarise the results of the BL Lac observations. No
significant excesses were detected from any of the objects on time
scales of days, months or years. The mean exposure on each object was
5.5 hours for each season that it was observed. Typically, in order to
detect a signal at the 4$\sigma$ level during a 5 hour exposure, the
object would need to have had a flux of at least 0.4 times that of the
Crab above the peak response energy for that season.

Flux upper limits were calculated for each object for each
season. Costamante \& Ghisellini [2] have made predictions for the TeV
flux from fourteen of the BL Lacs included in this paper using two
different methods. The upper limits presented here were compared with
these predictions; those of six BL Lacs, highlighted in Tables 1
\& 2, were found, during a number of observing seasons, to be lower
than the predicted fluxes according to the Fossati approach [4]
adapted in [2]. It should be noted however, that the upper limits
quoted here pertain only to the specific period during which the
observations were made. Indeed, 1ES2344, H1426 and 1ES1959 were
initially observed as part of this BL Lac campaign and, like the
objects listed here, were not detected. In subsequent years,
continuous monitoring, similar to that described here, revealed these
objects to be TeV emitters when in more active states.

The analysis of these results is ongoing. Spectral energy
distributions are being constructed and the flux predictions are being
corrected to account for pair-production of the gamma rays with the
infra red background radiation so that our upper limits can be
compared to these predictions in a more meaningful way.

\section{References}

\vspace{\baselineskip}

\re
1.\ Catanese M. A. \ et al. \ 1998, ApJ 501, 616
\re
2. Costamante L. \& Ghisellini G. \ 2002, A\&A 384, 56
\re
3.\ D'Vali M. \ et al. \ 1999, in Proc of 26th ICRC, Salt Lake City, 3, 422
\re
4.\ Fossati G. et al. \ 1998, MNRAS 299, 433
\re
5.\ Horan D. \ et al. \ 2001, AIP Conf. Proc. 587, 324
\re
6.\ Nishiyama T. \ et al. \ 2000, AIP Conf. Proc. 516, 26, 369
\re
7.\ Punch M. \ et al. \ 1992, Nature 358, 477
\re
8.\ Quinn J. \ et al. \ 1998, ApJ 456, L83
\re
9.\ Vernon-Cetty M. P. \& Vernon P. \ 1993, A\&AS 100, 521

\endofpaper
\end{document}